\documentclass[preprint,preprintnumbers,amsmath,amssymb,11pt]{revtex4}
\usepackage{amsmath,epsfig,graphicx,amssymb}

\begin{document}


\title{SKYRME-HARTREE-FOCK DESCRIPTION OF THE DIPOLE STRENGTH IN NEUTRON-RICH
TIN ISOTOPES}

\author{J. KVASIL$^{1}$, V. O. NESTERENKO$^{2}$, W. KLEINIG$^{2,3}$,
D. BO\v Z\' IK$^{1}$, and P.-G. REINHARD$^{4}$}

\affiliation{$^{1}$\it Institute of Particle and Nuclear Physics, Charles
University,
CZ-18000 Praha, Czech Republic\\
kvasil@ipnp.troja.mff.cuni.cz; bozik@ipnp.troja.mff.cuni.cz}

\affiliation{$^{2}$\it BLTP, Joint Institute for Nuclear Research, 141980,
Dubna,
Moscow region, Russia\\
nester@theor.jinr.ru; kleinig@theor.jinr.ru}

\affiliation{$^{3}$\it Technische Universit\"at Dresden, Inst. f\"ur Analysis,
D-01062, Dresden, Germany.}

\affiliation{$^{4}$\it Institut f\"ur Theoretische Physik II, Universit\"at
Erlangen, D-91058,
Erlangen, Germany;\\
mpt218@theorie2.physik.uni-erlangen.de}

\date{today}

\begin{abstract}
  Low-energy E1 strength in neutron-rich $^{132-164}$Sn isotopes is analyzed
  in the framework of the Skyrme random phase approximation (RPA) with
  different Skyrme forces. A double folding procedure is applied to take into
  account the energy-dependent width effects beyond RPA. All the considered
  Skyrme forces indicate a soft prolate
  deformation in the open shell isotopes $^{142-164}$Sn. The integrated
  E1 strength in the energy region of the pygmy resonance
  grows with the neutron number. The influence of
  deformation on the integrated strength near the particle emission
  thresholds (which is of a keen astrophysical interest) is strictly
  suppressed by the mutual compensation effect for the branches of the giant
  dipole resonance. The results obtained are in a good agreement with the
  previous findings of the relativistic mean field model.
\end{abstract}

\maketitle

\newcommand{\vek}[1]{\mathbf{#1}}
\newcommand{\me}{\mathrm{e}}
\newcommand{\mi}{\mathrm{i}}
\newcommand{\radi}{(\vek{r})}
\newcommand{\radii}{\vek{r}}
\newcommand{\der}[2]{\mathrm{d}^{#2} #1 \,}
\newcommand{\bra}[1]{\langle #1 |}
\newcommand{\ket}[1]{| #1 \rangle}
\newcommand{\squary}[1]{\left[ #1\right]}
\newcommand{\wavy}[1]{\left( #1\right)}

\section{Introduction}

Exotic nuclear collective excitations, like the dipole pygmy mode (PM),
represent a subject of intense investigations during last decades, see the
recent review \cite{Paar_Vretenar_2007}. The analysis of these excitations
provides a deeper insight into nuclear dynamics. Besides, the PM affects the E1
strength near the particle emission thresholds, which is of a key interest for
astrophysical applications \cite{Paar_Vretenar_2007,Stone_Reinhard_2007}. The
PM is viewed as a vibration of the neutron skin against the isospin-saturated
proton-neutron core \cite{Paar_Vretenar_2007}.  In this connection, the
evolution of PM with rising the neutron number and approaching the neutron drip
line attracts a large attention.

The available experimental data show the growth of PM strength with neutron
excess in medium-heavy \cite{Wagner_2008,Savran_2006,Savran_2008} and light
\cite{Anmann_1999,Tryggestad_2001} {\it stable} nuclei.  This trend was
confirmed by relativistic
and non-relativistic (Skyrme)
mean-field calculations \cite{Paar_Vretenar_2007}.
Meanwhile, the experimental effort focuses on drip-line regions and thus a
theoretical analysis of the PM behaviour in unstable nuclei is required.  Such
study was recently performed for neutron-rich $^{132-166}$Sn isotopes within
the relativistic mean-field model (RMF) \cite{Pena_Arteaga_2009}.  A growth
of PM strength with the neutron number was confirmed and a region of deformed
isotopes $^{142-162}$Sn with more spread of the strength was predicted.

The aim of present paper is to investigate the deformation and neutron-excess
effects in the $^{132-166}$Sn chain within the Skyrme mean-field approach. The
separable random-phase-approximation (SRPA) model
\cite{Nesterenko_2002,nest_PRC_06} is used. SRPA is fully self-consistent as
both the static mean field and residual interaction are derived from the same
Skyrme functional. The model covers both spherical and axially-deformed nuclei.
It was already applied to describe electric
\cite{Nesterenko_2002,nest_PRC_06,nest_ijmpe_07,nest_PRC_08} and magnetic
\cite{Ve09,Nesterenko_2010_2} giant resonances as well as the E1 strength near
the particle thresholds \cite{Kvasil_IJMPE_09}. In the present paper, the SRPA
formalism is supplemented by the double folding procedure which allows to
compute strength functions with an energy-dependent width. The SRPA analysis
below is done with the representative set of Skyrme forces,  SkT6 \cite{skt6},
SkM* \cite{skms}, SLy6 \cite{sly6}, and SkI3 \cite{ski3}, covering a broad
interval of Skyrme parameters and nuclear matter characteristics.

\section{SRPA approach}

SRPA is derived from the Skyrme energy functional \cite{Bender_2003}
\begin{equation}\label{eq:func}
    E(\rho, \tau, \vek{s}, \vek{J}, \vek{T}, \chi) = \int \mathcal{H}
    \radi d\vek{r}
\end{equation}
involving time-even (nucleon $\rho_q$, kinetic-energy $\tau_q$, spin-orbit
$\vek{J}_q$) and time-odd (current $\vek{j}_q$, spin  $\vek{s}_q$,
vector kinetic-energy $\vek{T}_q$) densities, as well as  the
pairing density $\chi_q$. The index $q$ stands for
neutrons ($q = n$) and protons ($q = p$).

The energy density $\mathcal{H} \radi$ consists of the kinetic-energy
$\mathcal{H}_{\mathrm{kin}} \radi$,
Skyrme $\mathcal{H}_{\mathrm{Sk}} \radi$, pairing $\mathcal{H}_{\mathrm{pair}} \radi$,
and Coulomb $\mathcal{H}_{\mathrm{Coul}} \radi$ contributions
\begin{equation}\label{eq:energdens}
    \mathcal{H} \radi = \mathcal{H}_{\mathrm{kin}}
    \radi+ \mathcal{H}_{Sk} \radi+ \mathcal{H}_{\mathrm{pair}}
    \radi+ \mathcal{H}_{\mathrm{Coul}} \radi
\end{equation}
where
\begin{eqnarray}
    \mathcal{H}_{\mathrm{kin}} \radi &=& \frac{\hbar^2}{2m} \tau \label{eq:Hkin} \; ,
    \\
    \mathcal{H}_{\mathrm{Sk}} \radi &=&
  \frac{b_0}{2} \rho^2- \frac{b'_0}{2} \sum_q\rho_{q}^2
  + \frac{b_3}{3} \rho^{\alpha+2}
  - \frac{b'_3}{3} \rho^{\alpha} \sum_q \rho^2_q
\nonumber
\\
 &&
 +b_1 (\rho \tau - \textbf{j}^2)
 - b'_1 \sum_q(\rho_q \tau_q - \textbf{j}^2_q)
 - \frac{b_2}{2} \rho\Delta \rho
 + \frac{b'_2}{2} \sum_q \rho_q \Delta \rho_q
\nonumber
\\
 &&
 - b_4 (\rho \nabla\textbf{J}\!+\!(\nabla\!\times\!\textbf{j})\!\!\cdot\!\!\textbf{s})
 - b'_4 \sum_q (\rho_q \nabla\textbf{J}_q\!
 +\!(\nabla\!\times\!\textbf{j}_q)\!\!\cdot\!\!\textbf{s}_q)
\nonumber
\\
  &&
  + \frac{\tilde{b}_0}{2} \textbf{s}^2
  - \frac{\tilde{b}'_0}{2} \sum_q \textbf{s}_{q}^2
+ \frac{\tilde{b}_3}{3} \rho^{\alpha} \textbf{s}^2
- \frac{\tilde{b}'_3}{3} \rho^{\alpha} \sum_q \textbf{s}^2_q
 -\frac{\tilde{b}_2}{2} \textbf{s} \!\cdot\!
  \Delta \textbf{s} + \frac{\tilde{b}'_2}{2}
  \sum_q \textbf{s}_q \!\cdot\!\Delta \textbf{s}_q
\nonumber
\\
 &&
  + \tilde{b}_1
   (\textbf{s}\!\cdot\!\textbf{T}\!-\!\textbf{J}^2)
  + \tilde{b}'_1
   \sum_q (\textbf{s}_q\!\cdot\!\textbf{T}_q
    \!-\!\textbf{J}_q^2) \; ,
    \label{eq:Skyrme}
    \\
    \mathcal{H}_{\mathrm{pair}} \radi &=& \frac{1}{4} \sum_{q} \chi_q^2 V_q
    \squary{1 - \wavy{\frac{\rho}{\rho_{0}}}^\gamma} \; ,
    \label{eq:Hpair}
    \\
    \mathcal{H}_{\mathrm{Coul}} \radi &=& \frac{e^2}{2} \int d\vek{r'} \rho_p \radi
    \frac{1}{|\radii - \radii'|} \rho_p (\radii') - \frac{3}{4} e^2
    \wavy{\frac{3}{\pi}}^{1/3} \squary{\rho_p \radi}^{3/4}\label{eq:HCoul}
\end{eqnarray}
where the densities without index, like $\rho = \rho_p + \rho_n$, are total.
The explicit expressions for the densities can be found elsewhere
\cite{nest_PRC_06}. In the part (\ref{eq:Hpair}), $V_q$ is the pairing strength
 and $\rho_{0}$ is the equilibrium nuclear matter density. The pairing
 is treated at the BCS level. The Coulomb contribution includes direct
 and exchange terms. In the part (\ref{eq:Skyrme}),
the Skyrme terms
with the tilded parameters $\tilde{b}_i$ and $\tilde{b}'_i$ are relevant
only for magnetic excitations and so are omitted in the present study.

The self-consistent derivation \cite{Nesterenko_2002,nest_PRC_06} gives
from (\ref{eq:func})-(\ref{eq:HCoul})  the SRPA Hamiltonian
\begin{equation}\label{eq:Hamiltonian_residual}
    \hat{H} = \hat{h}_{\mathrm{HFB}} + \hat{V}_{\mathrm{res}}
\end{equation}
where $\hat{h}_{\mathrm{HFB}}$ is the HFB mean field
\begin{equation}\label{eq:HFBmf}
    \hat{h}_{\mathrm{HFB}} = \int d{\textbf r} \sum_{\alpha_+}
    \squary{\frac{\partial E}{\partial J_{\alpha_+} \radi}} \hat{J}_{\alpha_+} \radi
\end{equation}
and $\hat{V}_{\mathrm{res}}$ is the separable  residual interaction
\begin{equation}\label{eq:residual_interaction}
    \hat{V}_{\mathrm res} = \frac{1}{2} \sum_{k, k' = 1}^{K} \squary{\kappa_{k k'}
    \hat{X}_k \hat{X}_{k'}+ \eta_{k k'} \hat{Y}_k \hat{Y}_{k'}}
\end{equation}
with one-body operators
\begin{eqnarray}\label{eq:operators}
    \hat{X}_k &=& i \int d\vek{r} \int d\vek{r'} \sum_{\alpha_+, \alpha'_+}
    \squary{\frac{\partial^2 E}{\partial J_{\alpha_+} (\radii')
    \partial J_{\alpha_+} \radi}} \bra{} \squary{\hat{P}_k, \hat{J}_{\alpha_+}
    \radi} \ket{} \hat{J}_{\alpha'_+} (\radii') \; ,
    \nonumber\\
    \hat{Y}_k &=& i \int d\vek{r} \int d\vek{r'} \sum_{\alpha_-, \alpha'_-}
    \squary{\frac{\partial^2 E}{\partial J_{\alpha_-} (\radii')
    \partial J_{\alpha_-} \radi}} \bra{} \squary{\hat{Q}_k, \hat{J}_{\alpha_-}
    \radi} \ket{} \hat{J}_{\alpha'_-} (\radii')
\end{eqnarray}
and inverse strength matrices
\begin{equation}
\label{eq:kappa_eta}
  \kappa_{k k'}^{-1 } =
  - i \langle [\hat{P}_{k},{\hat X}_{k'}] \rangle \; , \quad
  \eta_{k k'}^{-1 }
  = -i
  \langle [\hat{Q}_{k},{\hat Y}_{k'}] \rangle \; .
\end{equation}
Here $\alpha_+ = \rho, \tau, \vek{J}, \chi$ and $\alpha_- = \vek{j}, \vek{s},
\vek{T}$ enumerate time-even $J_{\alpha_{+}}$ and time-odd
$J_{\alpha_{-}}$densities; $\hat J_{\alpha_{\pm}}$ are the density operators;
$\hat{Q}_{k}$ and $\hat{P}_{k}=i[\hat H,\hat{Q}_{k}]$ are time-even and
time-odd hermitian input operators chosen following the procedure
\cite{nest_PRC_06,nest_PRC_08}. The operators of the residual interaction
$\hat{X}_k$ and $\hat{Y}_k$ are time-even and time-odd, respectively.
The single-particle Hamiltonian
$\hat{h}_{\mathrm{HFB}}$ is determined by the first functional derivatives of
the initial functional (\ref{eq:func}) while operators $\hat{X}_k$ and
$\hat{Y}_k$ are driven by the second functional derivatives of the same
functional. Hence the model is fully self-consistent. The number $K$ of
separable terms is determined by the number of the input operators
$\hat{Q}_{k}$. Usually we have $K=3 - 5$. Hence a low rank of the RPA
matrix and minor computational effort even for heavy deformed nuclei.

When studying giant resonances, one is usually interested in
the strength function. For electric dipole excitations, the strength function
reads
\begin{equation}
\label{eq:strength_function}
 S_{E1}(E)=\sum_{\mu} S_{E1\mu}(E), \quad
  S_{E1\mu}(E) = \sum_{\nu}
  E_{\nu}|\langle \nu|\hat{f}_{E1\mu}|\rangle|^2
  \zeta_{\Delta}(E - E_{\nu})
\end{equation}
where $\hat{f}_{E1\mu} = \sum_i e_i^{\mathrm{eff}} e r \mathrm{Y}_{1\mu}$
is operator of $E1\mu$-transition with the effective charges
$e_p^{\mathrm{eff}} = N/A$ and $e_n^{\mathrm{eff}} = -Z/A$ and
\begin{equation}
\label{eq:lorentz}
    \zeta_\Delta(E-E_\nu) = \frac{1}{2\pi}
    \frac{\Delta}{\wavy{E-E_\nu}^2 + \wavy{\frac{\Delta}{2}}^2}
\end{equation}
is the Lorentz weight with the averaging parameter $\Delta$ to simulate the
smoothing effects beyond SRPA (escape widths and coupling to complex
configurations). The strength function is composed by computing the
detailed RPA spectrum $E_{\nu}$ and reduced transition probabilities
$B(E1,\nu)=|\langle\nu|\hat{f}_{E1\mu}|\rangle|^2$. However it is more
convenient to compute the strength function directly by using the linear response
techniques \cite{Nesterenko_2002,nest_PRC_06} for which the factorization
(\ref{eq:residual_interaction}) additionally simplifies the calculation. We
thus use this direct technique.

The actual smoothing effects depend on the excitation energy $E$. The escape
widths appear above the particle emission thresholds and grow with energy due
to widening of the emission phase space. The coupling with complex
configurations (collisional width) also increases with energy. To take into
account these trends, one should use in
(\ref{eq:strength_function})-(\ref{eq:lorentz}) an energy-dependent averaging
parameter $\Delta (E)$. This can be done by implementation of the double
folding scheme which allows to use the strength function formalism (with
fixed width) for the RPA solutions.  At the first step, we calculate the strength
function (\ref{eq:strength_function}) with a small but fixed value of
$\Delta$. This gives the strength distribution $S_{E1\mu} (E')$ very close to
the actual RPA one but for the equidistant energy grid.  At the next step,
this strength is additionally folded by using the energy dependent averaging
$\Delta (E)$:
\begin{equation}\label{eq:averaging}
    S'_{E1}(E) = \int \der{E'}{} S_{E1}(E') \xi_{\Delta (E')} (E-E') \; .
\end{equation}
In the present study we use the simple linear energy dependence
\begin{equation}\label{eq:delta}
    \Delta (E') = \left\{ \begin{array}{ll}
\Delta_0 & \mbox{for } E'\leq E_{\mathrm{th}} \\
\Delta_0 + a (E' - E_{\mathrm{th}}) & \mbox{for } E' > E_{\mathrm{th}}
\end{array}\right.
\end{equation}
where $E_{\mathrm{th}}$ is the energy of the first emission threshold and
$\Delta_0$ is a minimal width describing the coupling with
complex configurations below the threshold. The rate $a$ is
chosen to give a proper reproduction of the form of the giant dipole
resonance (GDR). In the present study, we use
$\Delta_0 = 0.1$ MeV and $a = 1/6$.

The  dipole  photoabsorption cross section
is related to the strength function (\ref{eq:strength_function}) as
$\sigma(E) \approx 0.402 S_{E1}(E)$ with $S_{E1}(E)$ being
in $e^2 \mathrm{fm}^2$ and $\sigma (E)$ in $\mathrm{fm}^2$.

\section{Results and discussion}

The results are obtained  for the representative set of Skyrme forces: SkT6
\cite{skt6},  SkM* \cite{skms}, SLy6 \cite{sly6}, and SkI3 \cite{ski3}.  These
forces have different nuclear matter characteristics and are widely used for
the description of ground state properties and dynamics of atomic nuclei
\cite{Bender_2003}, including deformed ones. The calculations use a cylindrical
coordinate-space grid with the mesh size 0.7 fm. The equilibrium quadrupole
deformations are found by minimization of the total energy of the system.
\begin{figure}[th] \label{Fig1}
\centerline{\psfig{file=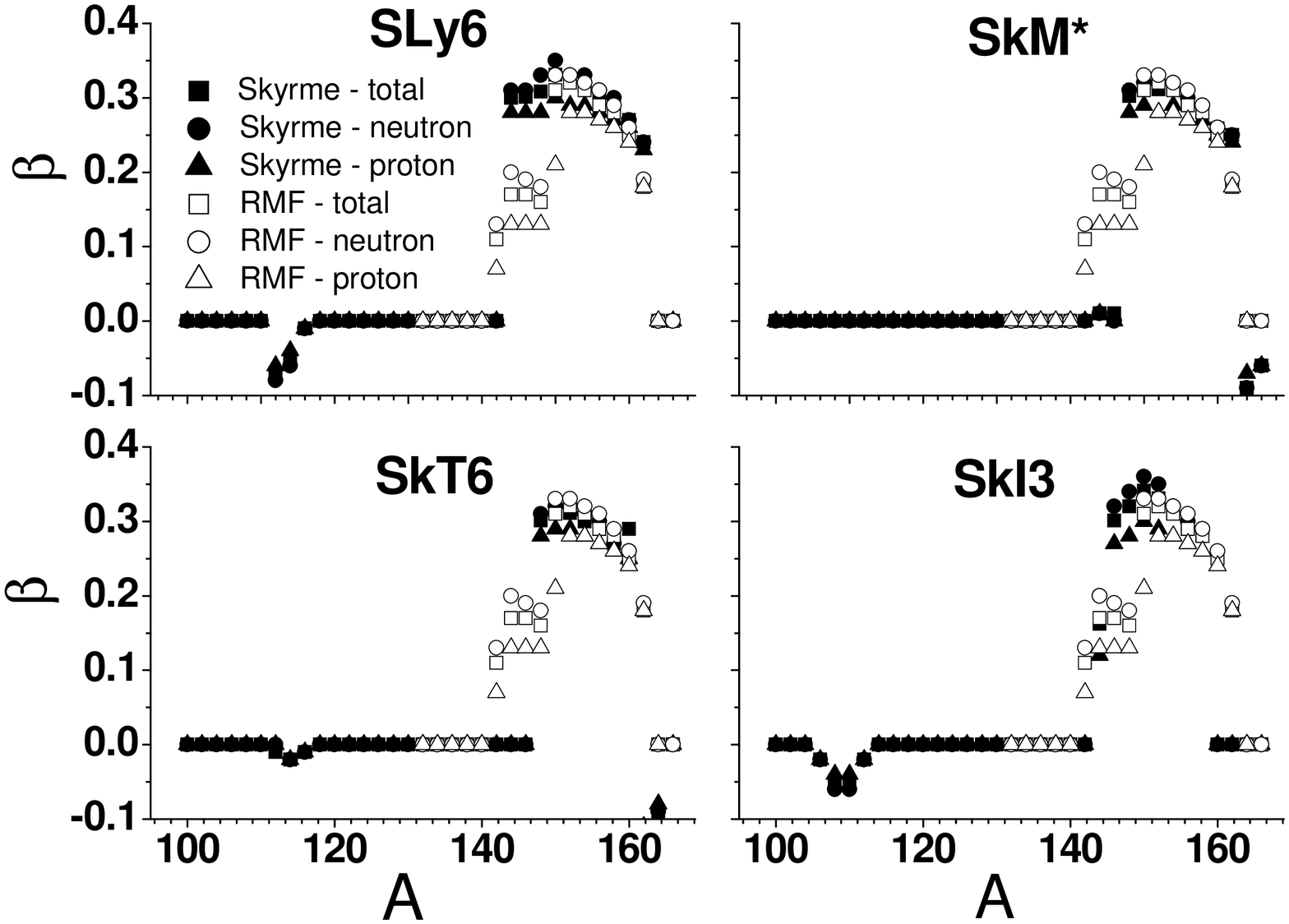,width = 7cm,angle=-90}} \caption{Equilibrium
total, neutron and proton quadrupole deformations in $^{100-166}$Sn, calculated
with Skyrme forces SLy6, SkM*, SkT6, and SkI3 and compared to the RMF results
\protect\cite{Pena_Arteaga_2009}.}
\end{figure}
%
\begin{figure}[t] \label{Fig2}
\centerline{\psfig{file=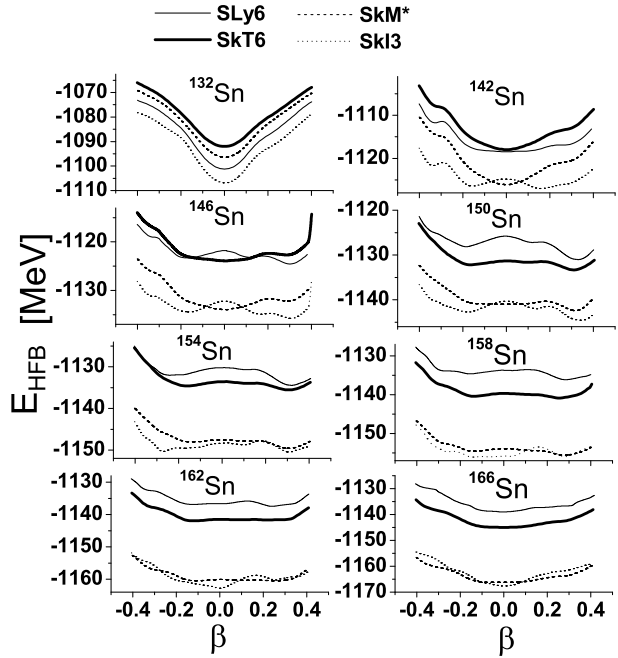,width = 9cm}} \caption{Dependence of the HFB
nuclear energy on the total quadrupole deformation $\beta$ for SkT6, SLy6,
SkM*, and SkI3 forces in selected neutron-rich Sn isotopes.}
\end{figure}

In Fig.1 the calculated equilibrium quadrupole deformations $\beta$ in Sn
isotopes with 100 $\le A \le 166$ are shown for the four different Skyrme
forces.  The broad mass region involves neutron-deficient, stable, and
neutron-rich nuclei. The deformations are given for the total,
neutron, and proton densities. The results are compared with the RMF findings
\cite{Pena_Arteaga_2009}. It is seen that for $A \le 140$ the isotopes keep
the spherical shape, for exception of a narrow region $112 \le A \le 116$
where forces SkT6, SLy6, and SkI3 predict some oblate softness. It is
remarkable that, despite the semi-magic character of Sn isotopes (Z=50), they
exhibit a significant prolate deformation in the neutron-rich region $142 \le
A \le 162$, which is in accordance to RMF predictions
\cite{Pena_Arteaga_2009}.  Note that neutron deformation in this region is
systematically larger than the proton ones, which may signify the higher
deformation of the excess neutrons.  Perhaps, just the drive of the large
neutron skin to the deformation causes the whole effect in these semi-magic
nuclei which otherwise might be spherical as in the neutron-deficit region.

The further analysis shows that $^{142-162}$Sn isotopes should be
rather considered as prolate-soft instead of prolate-deformed. Indeed, the
total HFB energies $E_{\mathrm{HFB}}$ depicted in Fig. 2 demonstrate for $142
\le A \le 162$ very shallow deformation minima.  Though the deformations and
$E_{\mathrm{HFB}}$ values in the transition regions with A=142-146
and 158-162 noticeably vary with the force (in some cases, e. g. for SkI3 in
$^{154}$Sn, both prolate and oblate minima are predicted), the results for
different forces are qualitatively similar. Altogether they point out to
possible strong softness and anharmonicity in $^{142-162}$Sn.
\begin{figure}[t] \label{Fig3}
\centerline{\psfig{file=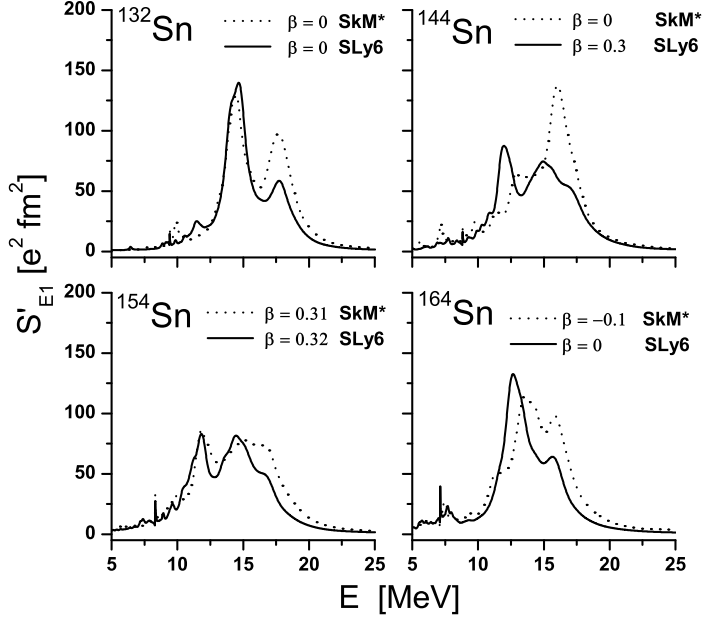,width = 10cm}} \caption{The doubly folded E1
strength function for SkM* and SLy6  in selected Sn isotopes.}
\end{figure}

In Fig. 3, the doubly folded strength function $S'_{E1}(E)$ is shown for the
forces SkM* and SLy6 in the representative isotopes (spherical $^{132}$Sn,
transition $^{144}$Sn, deformed $^{154}$Sn, and spherical $^{164}$Sn) covering
the deformation region and its vicinities. The double folding is computed with
the threshold energies (associated to the neutron single-particle Fermi levels)
marked in Fig. 4. It gives a reasonable width $\Delta \sim 1$ MeV at the GDR
peak. Due to larger averaging at higher energies, the double folding
considerably suppress the artificial strong shoulder at the right GDR side,
which often appears in Skyrme-RPA calculations both with
\cite{Nesterenko_2002,nest_ijmpe_07,nest_PRC_08} and without \cite{Mar_PRC_05}
the separable prescription. Fig. 3 shows that SkM* and SLy6 give similar shapes
for spherical and deformed isotopes but rather different deformations in
transition $^{144}$Sn. Some fine structures are seen in the pygmy energy region
5-10 MeV. The E1 strength in this region grows with the neutron number.

The latter effect is better visible in Fig. 4 where the E1 strength function
(\ref{eq:strength_function}) is exhibited with a small constant averaging
$\Delta = 0.1$ MeV in the low-energy pygmy region. It is seen that fine
structure of the strength significantly depends on the force and it is more
separated from the GDR in spherical isotopes than in deformed ones. Hence, in
accordance to RMF results \cite{Pena_Arteaga_2009}, spherical nuclei are
indeed more suitable for observation of the pygmy mode.  Note also that pygmy
energy regions stay near the particle emission thresholds and so the pygmy
resonance may affect the near-threshold E1 strength of the astrophysical
interest \cite{Paar_Vretenar_2007,Stone_Reinhard_2007,Kvasil_IJMPE_09}.

\begin{figure}[t] \label{Fig4}
\centerline{\psfig{file=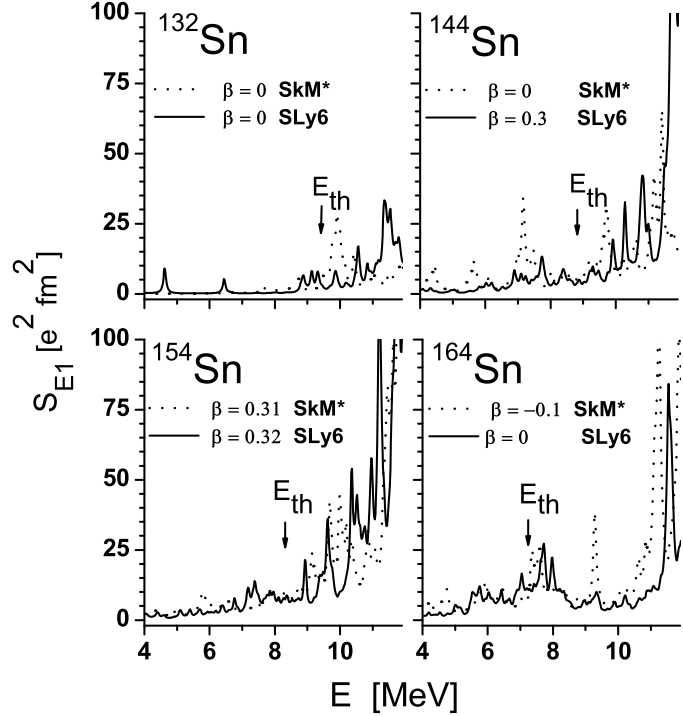,width = 9cm}} \caption{The low-energy E1
strength function (with constant $\Delta = 0.1$ MeV) for SkM* and SLy6 in
selected Sn isotopes. The estimated particle emission thresholds
$E_{\mathrm{th}}$ are marked by arrows.}
\end{figure}
%
\begin{figure}[th] \label{Fig5}
\centerline{\psfig{file=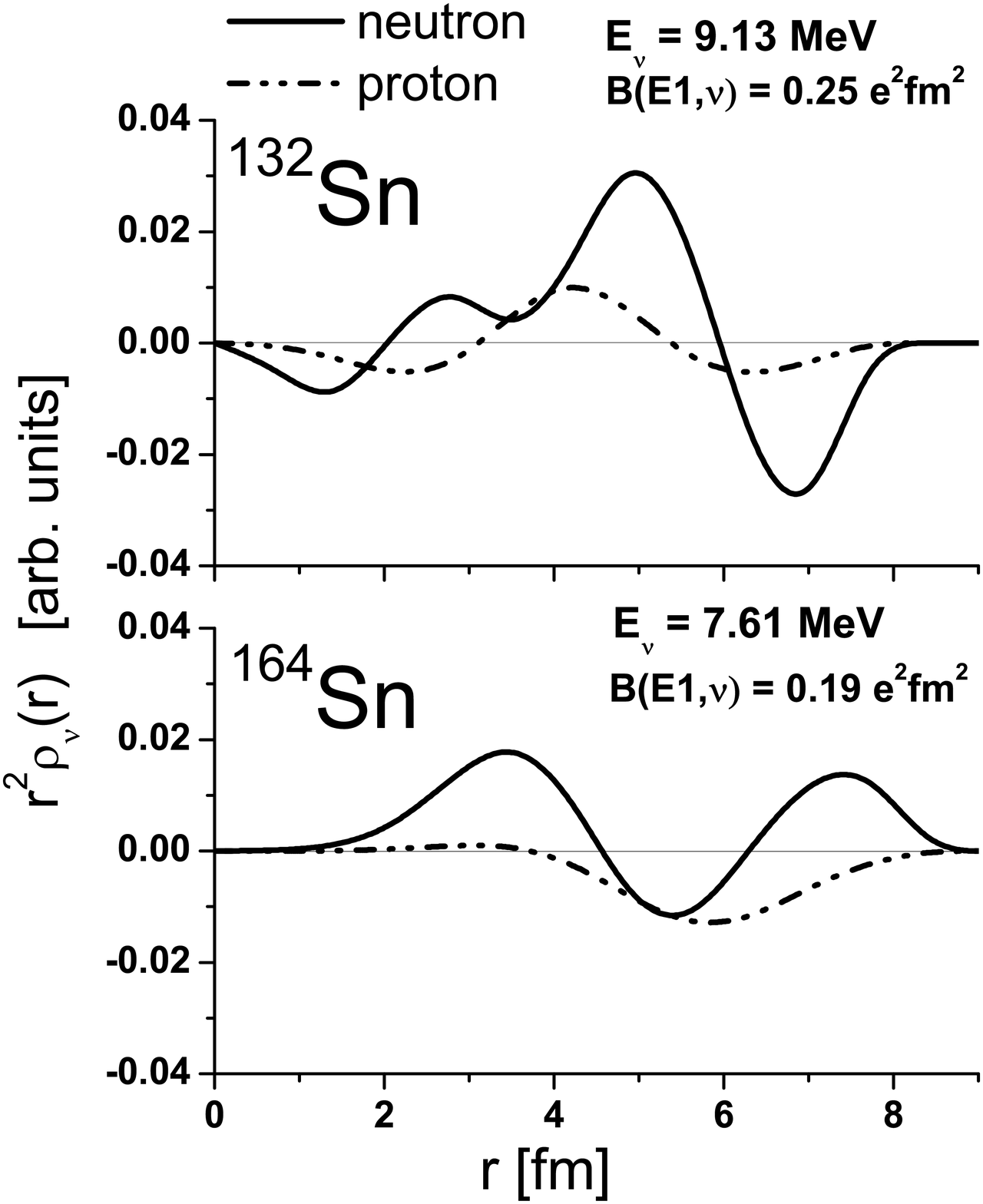,width = 6.5cm}} \caption{Neutron and proton
transition densities $\rho_{\nu} (r)$ in particular low-energy RPA $\nu$-states
with large $B(E1,\nu)$ values in spherical $^{132}$Sn and $^{164}$Sn. The force
SLy6 is used.}
\end{figure}
%
%
\begin{figure}[th] \label{Fig6}
\centerline{\psfig{file=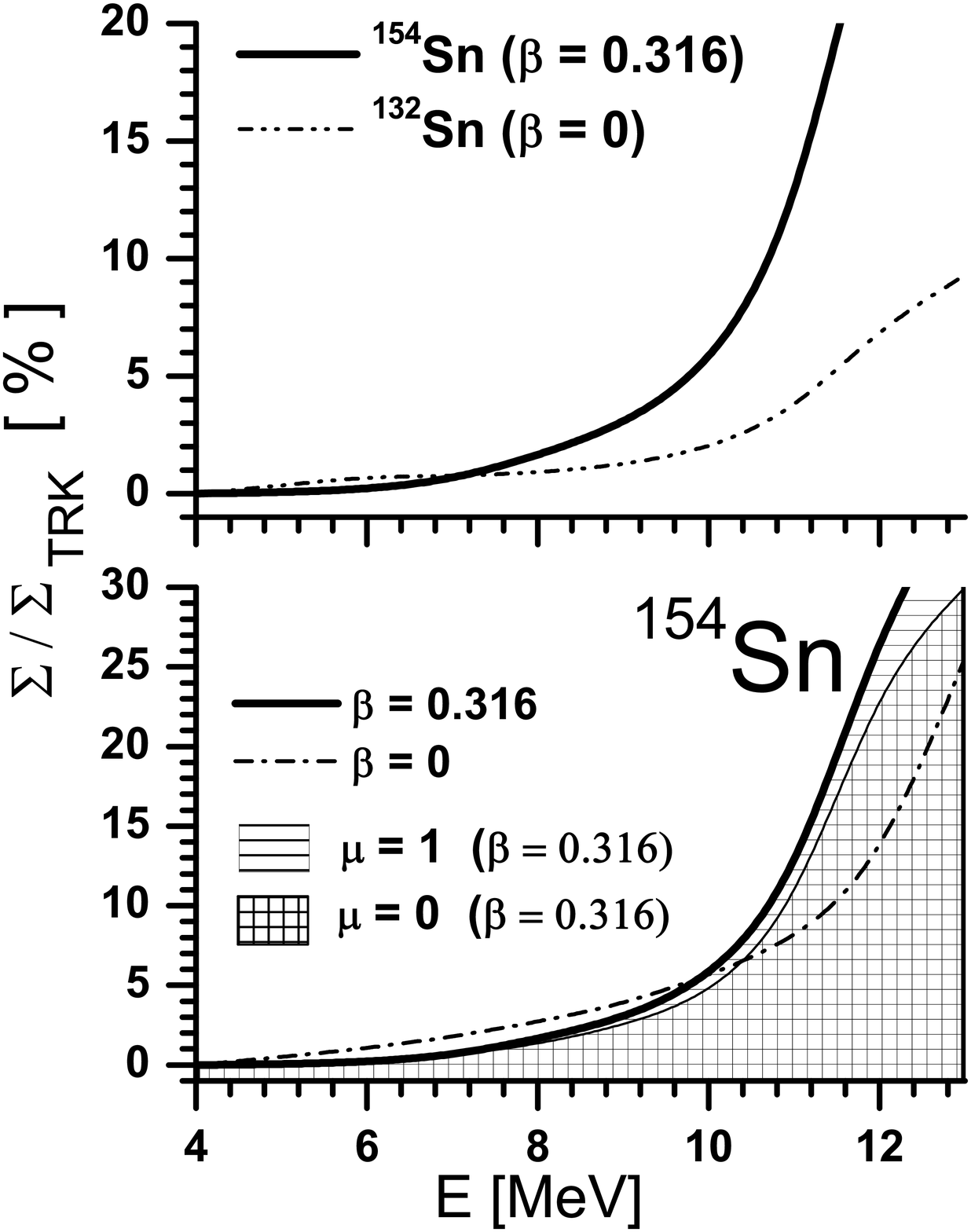,width = 6.5cm}} \caption{The integrated
photoabsorption cross section (\ref{eq:cumulative_cross_section}) in the
percentage of the TRK sum rule $\Sigma_{\mathrm{TRK}}$. The upper panel
compares the cross sections for spherical $^{132}$Sn and deformed $^{154}$Sn.
The bottom panel shows the cross sections in $^{154}$Sn for the equilibrium
deformation $\beta = 0.316$ and constrained spherical case $\beta = 0$. The
contributions of $\mu = 0$ and $\mu = 1$ GDR branches for $\beta = 0.316$ are
exhibited by areas with linear and check filling, respectively. The force SLy6
is used.}
\end{figure}

In Fig. 5, the pygmy nature of the low-energy strength is illustrated
for particular RPA states in spherical $^{132}$Sn and $^{164}$Sn.
It is seen that in the nuclear surface at 6-8 fm the neutron transition
density dominates thus manifesting the oscillation of the neutron excess
against the proton-neutron core.

Finally Fig. 6 shows the integrated E1 photoabsorption cross section
\begin{equation}\label{eq:cumulative_cross_section}
    \Sigma (E) = \int_{4 \mathrm{MeV}}^E \sigma(E') \der{E'}{}
\end{equation}
in units of the Thomas-Reiche-Kuhn (TRK) sum rule $\sum_{TRK} = 60NZ/A$ mb
MeV.  The upper panel compares the ratios in spherical $^{132}$Sn and deformed
$^{154}$Sn.  At $E >$ 7 MeV the isotope $^{154}$Sn delivers more low-energy E1
strength. This may be caused by two factors in $^{154}$Sn, the deformation and
the larger mass number. The deformation down-shifts the left $\mu$=0 GDR
branch and the larger nuclear size down-shifts the GDR in general. While the latter
factor is clear, the effect of deformation deserves more analysis
\cite{Kvasil_IJMPE_09,Donau_2007,Rusev_PRC_2009}. It is inspected for
$^{154}$Sn in the lower panel of Fig. 6. The comparison of the photoabsorption
cross sections for equilibrium deformation $\beta = 0.316$ and
deliberately constrained spherical case $\beta = 0$ shows that the deformation
results in a larger cumulative strength only at $E >$ 10 MeV, i.e. in a close
vicinity to the GDR. Instead, at the lower energies near thresholds
(the region of the astrophysical interest) the deformation leads, in fact, to
a weak depletion of the strength.

The weak deformation effect in the pygmy region can be explained by the
destructive competition of $\mu=0$ and $\mu=1$ GDR contributions to the
low-energy E1 strength \cite{Kvasil_IJMPE_09}.  As seen from Fig. 6, both
these GDR branches affect the pygmy region and the $\mu=0$ branch evidently
dominates. However, at low energies far enough from the GDR ($E <$ 10 MeV in
our case), the influences of both branches become more comparable. Indeed, the
$\mu=0$ branch, being closer to the pygmy region, carries at the same time
twice smaller strength than the $\mu=1$ one.  Hence the comparable effects of
the branches at the remote energies.  The deformation shifts the $\mu=0$ and
$\mu=1$ branches in the opposite directions, thus leading to the deformation
contributions of the opposite sign and comparable magnitude. These
contributions compensate each other and drastically suppress the total
deformation effect at the remote low energies.  In principle, the rest
effect can be of any sign, e.g. negative, as in our case. A weak negative
deformation effect agrees with findings \cite{Pena_Arteaga_2009} and
contradicts the assertion \cite{Donau_2007,Rusev_PRC_2009} on its strong and
positive magnitude.

\section{Conclusions}

The long chain of neutron-rich Sn isotopes with $132 \le A \le 166$ is
investigated within the Skyrme RPA method by using SkT6, SkM*, SLy6, and SkI3
Skyrme forces. Both spherical and axially deformed shapes are considered. The
smoothing effects beyond RPA are simulated by the double folding procedure
which allows a direct computation of the strength function with the energy
dependent width.

All the Skyrme forces predict a region of prolate-soft isotopes $^{142-162}$Sn
for which a strong anharmonicity in low-energy collective modes is
expected. Though prolate shapes look surprising for the semi-magic Sn
isotopes, they can be caused by the large neutron excess driving the
system to deformations.  The calculations also show that the pygmy E1
strength increases with the neutron number. Following our analysis, the nuclear
deformation does not noticeably affect this trend and leads mainly to more
spread and redistribution of the E1 strength. The effect of deformation on the
cumulative E1 strength near the particle emission thresholds is shown to be
small and negative (slight depletion of the strength). The result is explained
by the compensation of the deformation contributions from $\mu=0$ and $\mu=1$
GDR branches. This finding may be of a significant interest for astrophysical
applications.

The four Skyrme forces yield qualitatively similar results
though their numerical predictions can vary
for the total nuclear energy and fine structure of the pygmy mode.
Our results are in a good agreement with the RMF study
\cite{Pena_Arteaga_2009}

\section*{Acknowledgments}

The work was partly supported by the grants DFG RE-322/12-1, Heisenberg-Landau
(Germany - BLTP JINR), and Votruba - Blokhintsev (Czech Republic - BLTP JINR).
W.K. and P.-G.R. are grateful for the BMBF support under contracts 06 DD 9052D
and 06 ER 9063. Being a part of the research plan MSM 0021620859 (Ministry of
Education of the Czech Republic) this work was also funded by Czech grant
agency (grant No. 202/06/0363).

\end{document}